# Band structure of Si/Ge core-shell nanowires along [110] direction modulated by external uniaxial strain

Xihong Peng,[1*] Fu Tang,[2] Paul Logan[2]

[1]Department of Applied Sciences and Mathematics, Arizona State University, Mesa, AZ 85212

[2]Department of Physics, Arizona State University, Tempe, AZ 85287

## ABSTRACT

Strain modulated electronic properties of Si/Ge core-shell nanowires along [110] direction were reported based on first principles density-functional theory calculations. Particularly, the energy dispersion relationship of the conduction/valence band was explored in detail. At the Γ point, the energy levels of both bands are significantly altered by applied uniaxial strain, which results in an evident change of band gap. In contrast, for the K vectors far away from Γ, the variation of the conduction/valence band with strain is much reduced. In addition, with a sufficient tensile strain (~1%), the valence band edge (VBE) shifts away from Γ, which indicates that the band gap of the Si/Ge core-shell nanowires experiences a transition from direct to indirect. Our studies further showed that effective masses of charge carriers can be also tuned by the external uniaxial strain. The effective mass of the hole increases dramatically with a tensile strain, while strain shows a minimal effect on tuning the effective mass of the electron. Finally, the relation between strain and the conduction/valence band edge is discussed thoroughly in terms of site-projected wave-function characters.

* Corresponding author.  E-mail: xihong.peng@asu.edu.

1. **Introduction**

One-dimensional semiconductor nanostructures, such as Si and Ge nanowires, have attracted extensive research efforts over the past decades [1-12]. They are expected to play important roles as both interconnects and functional components in future nanoscale electronic and optical devices, such as light-emitting diodes (LEDs) [2, 3], ballistic field-effect transistors (FETs) [5, 6], inverters [3], and nanoscale sensors [4, 9]. Experimental and theoretical investigations showed that in these nanoscale wires the charge carriers are confined in the lateral direction of the wires, thus quantum confinement effects are expected to play an important role on the electronic properties. This confinement effects have been observed, for example, in photoluminescence studies, by exhibiting a substantial blue-shift in the emission spectrum with a reduction in the diameter of nanowires [13-15]. Researchers also found that the band gap of Si and Ge nanowires depends on several factors, such as size [14-16], crystalline orientation [15-17], surface chemistry [17, 18], and doping [17, 19].

Recently, a particular attention has been given to Si/Ge core-shell nanowires, in which factors, such as heterostructure composition and interface geometry, can be further manipulated to tune the electronic properties of the nanowires [18, 20-31]. Compared to the single composition of Si or Ge nanowires, the core-shell structure has some superior properties. For instance, a better conductance and higher mobility of charge carries can be obtained, due to the band offsets in the core-shell nanowires [18, 23, 30]. In addition to experimental studies [27, 31], several theoretical calculations were performed to study the quantum confinement effects in the Si/Ge core-shell nanowires [25, 28-30]. In these calculations, the band gap and near-gap electronic states are particularly investigated as a function of the hetero-composition. For example, Wang's group [28, 29] reported the band gap of the Si/Ge core-shell nanowires as a function of the composition with a diameter of the wires up to 3 nm; Migas *et al.* [32] studied the electronic properties of the Si/Ge core-shell nanowire along the [100] direction with a diameter of 1.5 nm; Yang *et al.* [30] investigated the near-gap electronic states with the core and shell regions along the [110] and [111] directions with a diameter up to 4 nm.

Strain is another factor that has been demonstrated to critically affect the electronic properties of various nanostructures. Despite a few limited experimental studies [23, 24], a detailed theoretical investigation of the strain effects on the electronic properties in Si/Ge core-shell nanowires is still lacking. In the Si/Ge core-shell nanowires, there is an intrinsic strain in the core-shell interface, due to the lattice mismatch between Si and Ge. Recently, Peng *et al.* [33] have shown that the band gap of the core/shell nanowires can be significantly modulated by the intrinsic strain. In present paper, we reported a thorough study of the effects of the external uniaxial strain on the band structures of Si/Ge core-shell nanowires with a diameter of the wires up to 5 nm, using first principles calculations.

## 2. Simulation details

Density-functional theory (DFT) [34] calculations were performed using VASP code [35, 36]. The local density approximation (LDA) was applied. In detail, a pseudo-potential plane wave approach was used with a kinetic energy cutoff of 200.0 eV. Core electrons were described using ultra-soft Vanderbilt pseudo-potentials [37]. The reciprocal space was sampled at $1 \times 1 \times 4$ using Monkhorst Pack meshes. A larger energy cutoff of 350.0 eV and a K-points mesh of $1 \times 1 \times 9$ were used to check the case of 2.5 nm Si-core/Ge-shell nanowire. No significant difference was found in the results using those parameters and the current setting. 21 K-points were included in the band structure calculations along the reciprocal direction $\Gamma$ to X. The dangling bonds on the wire surface were saturated by hydrogen atoms with the initial bond lengths 1.47 Å (Si-H), and 1.51 Å (Ge-H), respectively. Those Si-H and Ge-H bonds are allowed to relax during geometry optimization. For the Si-core/Ge-shell nanowires, the core is consisted of 30 Si atoms, and the thickness of the Ge shell varies (See Figure 1). Similarly, for the Ge-core/Si-shell wires, the core contains 30 Ge atoms and the size of the Si shell varies. The initial lattice constant along the [110] direction in the Si-core/Ge-shell nanowires was set to be 0.3977 nm, taken from the lattice constant of bulk Ge 0.5625 nm (*i.e.* $a_{initial[110]} = a_{bulk}/\sqrt{2}$). The lattice constant along the axial direction in the Ge-core/Si-shell nanowires was initially set to be 0.3862 nm, derived from the lattice constant of bulk Si 0.5461 nm. The reason to choose the different initial lattice constants is the following. Take an example of the Si-core/Ge-shell nanowires, the core consists of 30 Si atoms, while the thickness of the Ge shell varies. Therefore, the larger the diameter of the wire, the more Ge atoms the wire has. The optimized lattice constant is thus expected to be closer to that of bulk Ge. Besides the axial lattice constant, the lateral size of the simulation cell was chosen so that the distance between the wire and its replica (due to periodic boundary conditions) is more than 1.0 nm to minimize the interaction. The [110] axial lattice constant of all the core/shell wires was then optimized through the technique of total energy minimization till the forces acting on atoms are less than 0.02 eV/Å. The electronic properties of a wire, such as band gap and effective masses of charge carriers, were then calculated by solving the Kohn-Sham equation within the frame of DFT. The band gap is defined by the energy difference between the conduction band edge (CBE) and the valance band edge (VBE). The effective masses of the electron and hole can be readily calculated according to the formula $m^* = \hbar^2/(d^2E/dk^2)$ from the band structure.

Table 1 lists the Si/Ge core-shell nanowires studied in present work. D is the diameter of a wire in the unit of nanometers, defined as the longest distance between two Ge atoms (for a Si-core/Ge-shell wire) or two Si atoms (for a Ge-core/Si-shell wire) in the cross-section; $N_{core}/N_{shell}$ are the numbers of core/shell atoms in a given wire; $N_H$ represents the number of H atoms needed to passivate the surface dangling bonds. Figure 1 gives the snapshots of three core/shell nanowires with the diameters of 2.5 nm, 3.7 nm, and 4.7 nm, viewed from the cross-section. The core and shell atoms could be either Si or Ge. The diameters of the core and the entire nanowire are also given at the bottom of the figure.

Once the geometrically optimized wire is obtained, uniaxial strain was applied by rescaling the axial lattice constant of the nanowire. For instance, a tensile strain of 2% means the axial lattice and the z coordinates of the atoms was rescaled to 102% of their original values, while a compressive strain of 2% implies the axial lattice and the z coordinates was rescaled to 98% of their original values. The positive values of strain refer to uniaxial expansion, while negative corresponds to compression. For each strained wire, the lateral *x* and *y* coordinates are further relaxed through energy minimization. Our study showed that the band structure of a wire is significantly modulated by strain.

## 3. Results and discussion

The optimized lattice constants for the Si/Ge core-shell nanowires are reported in Table 2. In the Si-core/Ge-shell wires, the lattice constant increases with the diameter of a wire, from 0.3917 nm for the 2.5 nm wire to 0.3944 nm for the 4.7 nm wire. In addition, the lattice constants for the Si-core/Ge-shell wires are smaller than 0.3977 nm (from bulk Ge), but larger than 0.3862 nm (from bulk Si). In contrast, the lattice constant of the Ge-core/Si-shell wires reduces with size, from 0.3985 nm to 0.3900 nm. The result is expected since a lager Ge-core/Si-shell wire has more Si atoms. With the optimized lattice constants, an intrinsic strain has been produced in the wires. The Si composition is in an intrinsic tensile strain, while the Ge composition experiences an intrinsic compressive strain [33].

The calculated band structures for the Si/Ge core-shell nanowires with different sizes are presented in Fig. 2(a) - 2(h). As an example, the band structure of 2.5 nm Ge-core/Si-shell nanowires under strain is plotted in Fig. 2(i) - 2(m). Detailed analysis of the band structures, such as band gap and effective masses of charge carriers, are shown below.

### I. Electron charge distribution of conduction and valence bands

Figure 3 shows the iso-value surfaces of the charge density of the VBE and CBE of the nanowires with the diameters 2.5 nm and 3.7 nm. From Figure 3(a) and 3(b), the charge of the VBE in the Si-core/Ge-shell wires is mainly distributed in the Ge shell, while the charge of the CBE is mainly located in the Si core. On the other hand, the charge of the VBE in the Ge-core/Si-shell wires is primarily in the Ge core, while the charge of the CBE is distributed in the Si shell, shown in Figure 3(c) and 3(d). It is concluded that the VBE charge is primarily in the Ge atoms while the CBE charge is in the Si atoms, regardless of the core-shell composition [30], which is a typical type II band alignment. This confined distribution of the electron charge results in a reduced band gap compared to that of the single composition Si or Ge nanowires at a given diameter [33].

### II. Band gaps

Bulk Si and Ge are materials with an indirect band gap. However, the Si/Ge core-shell, the single composition Si and Ge nanowires along the [110] direction demonstrate a direct band gap at Γ [14, 16, 17]. The DFT predicted band gaps for the Si/Ge core-shell wires are reported in Table 2. The gap increases when the size of a wire reduces, mainly due to the quantum confinement effects. It is known that DFT underestimates band gap of semiconductors, and advanced GW method [38-40] can provide improved predictions. However, for the size of the nanowires investigated in present work, GW is not applicable due to its extremely high computing cost. The present work is mainly focused on the variation of electronic properties under factors such as external strain and size. Previous studies [41] on Si nanoclusters showed that the energy gap calculated by DFT obeys a similar strain-dependency as the optical gap predicted by advanced configuration interaction (CI) method and the quasi-particle gap (defined as the difference of ionization potential and electron affinity). In addition, DFT gap predicts a similar size-dependency as the optical gap obtained using GW and quantum Monte Carlo methods [42, 43].

Our calculations showed that the band structure and band gap of the Si/Ge core-shell nanowires can be modulated by strain. The variations of the band gap with strain for both Si-core/Ge-shell and Ge-core/Si-shell nanowires are plotted in Figure 4. Generally, the gap decreases evidently with tensile strain, while it only slightly varies with compression. Interestingly, under a sufficient tensile strain (~ 1%), the nanowires with a large diameter (> 3 nm) demonstrate an indirect band gap, indicated by the green arrows in Figure 4. This transition was discussed in detail in next section.

To understand the general trend in Figure 4 (the gap reduces with the tensile strain), the energies of the CBE and VBE were further examined. As an example, Figure 5 shows the energy variations of the CBE and VBE with strain in the 2.5 nm Si-core/Ge-shell nanowire. Note that a similar behavior was also observed for other wires. From Figure 5, the energies of the CBE and VBE both decrease with the tensile strain and the CBE curve has a lightly larger slope compared to that of the VBE curve, which may be understandable from a detailed analysis of the wave-function character of the CBE and VBE. The wave-function of various electronic states including CBE and VBE has been projected onto spherical harmonics within spheres of a radius of 1.2 Å around each Si and Ge ion. The decomposed contributions/coefficients of *s*, *p*, *d* orbitals are listed in Table 3. The VBE is dominated by a $p_z$ character, which suggests that the nodal surfaces of the positive and negative values of the wave-function are perpendicular to the axis of the wire [44, 45]. Under a tensile strain, the distance between the nodal surfaces increases. And the kinetic energy associated with the electron transportation between atoms reduces [45]. So does the VBE energy. For the CBE, the projected wave-function includes a significant $d_z 2$ character, which is consistent with literature [46-48] that the *d* character defines the orbitals of the conduction band in a crystal. Similar to the $p_z$ character, the $d_z 2$ orbit also produces the nodal surfaces perpendicular to the axis of the wire. Therefore, the CBE energy reduces when the nanowire is under a tensile strain.

To understand a larger slope in the CBE curve compared the VBE curve, a strain analysis in the core and shell regions is shown. For the relaxed 2.5 nm Si-core/Ge-shell nanowire, the intrinsic strain in the Si-core and the Ge-shell are +1.5% and -1.5%, respectively [33]. If a +2% external tensile strain is applied, the resulting strain in the Si-core and Ge-shell are approximately 3.5% and 0.5%, respectively. Thus the Si atoms in the core are more effectively expanded compared to the Ge atoms in the shell. Since the charge of the CBE is mainly located in the Si-core while the charge of the VBE is in the Ge-shell, the energy of the CBE shows a larger reduction compared to that of VBE when the wire is under external tensile strain.

### III. Energy dispersive curves and effective masses

The effective masses of the electron and hole are reported in Table 2. $m_e^*$ represents the effective mass of the electron, while $m_h^*$ is the effective mass of the hole, in unit of free electron mass $m_e$. The effective mass of the electron are 0.13 $m_e$ or 0.14 $m_e$, having a negligible change with size and the core-shell composition. In contrast, the effective mass of the hole is dependent on the wire size and composition.

The strain effect on the effective masses of the electron and hole were further studied. Taking the 2.5 nm Ge-core/Si-shell wire as an example, the dispersion relation with the K vector in the range of $\pm 0.2 \cdot \frac{2\pi}{a}$ is plotted under different values of strain in Figure 6(a). It shows that the strain has a dominant effect on the band structure at Γ - the energies are shifted evidently. However, the strain has a negligible effect on the bands with K vectors away from Γ (K > $0.15 \cdot \frac{2\pi}{a}$ or K < $-0.15 \cdot \frac{2\pi}{a}$). These strain effects may be understandable from a simple tight-binding model discussed in the reference [49]. For an in-depth understanding, a detailed analysis of the wave-functions is reported in Table 3. For the valence/conduction band at Γ, there is a significant portion of $p_z/d_z^2$ character. Therefore, the energy of the valence/conduction band at Γ decreases with an external tensile strain. However, for the wave-functions at K = $0.20 \cdot \frac{2\pi}{a}$, the portion of $p_z/d_z^2$ character is largely reduced, while the $p_x/p_y$ character dominates. In this case, the nodal surfaces of the wave-functions are parallel to the axis of the nanowire and the distance between the nodal surfaces is negligibly changed by a uniaxial strain. Therefore, the kinetic energy associated with the electron transportation stays the same, giving a minimal energy shift with strain [45].

From Figure 6(a), the CBE is located at $\Gamma$, regardless of strain. However, the VBE shows an interesting transition – it is no longer located at $\Gamma$ for a large tensile strain (~ 1.8%), implying an indirect band gap. This direct-to-indirect gap transition with strain is clearly demonstrated in Fig. 6. In Fig. 6(a), the energy of the valence band at $\Gamma$ (K = 0.0) is noticeably decreased with tensile strain, where the energy shift of the valence band at K vectors away from $\Gamma$ (i.e. $K > 0.05 \cdot \frac{2\pi}{a}$) is negligible with strain. Therefore, with a sufficient tensile strain, the energy of the valence band at $\Gamma$ can be reduced to an extent so that it is lower than the energy at $K = 0.05 \cdot \frac{2\pi}{a}$. An enlarged graph of the valence band under +1.8% strain is presented in Figure 6(b). The nature of the band gap (direct or indirect) is determined by the energies of the two states labeled as $v_0$ at $\Gamma$ and $v_1$ at $K = 0.05 \cdot \frac{2\pi}{a}$. Without strain, the energy of $v_0$ is higher than that of $v_1$ (refer to Figure 6(a)), indicating a direct band gap. With 1.8% tensile strain, the energy of $v_1$ is higher than that of $v_0$, giving an indirect band gap. This direct-to-indirect gap transition with strain is clearly demonstrated in Fig. 6. In Fig. 6(a), the energy of the valence band at $\Gamma$ (K = 0.0) is noticeably decreased with tensile strain, where the energy shift of the valence band at K vectors away from $\Gamma$ (i.e. $K > 0.05 \cdot \frac{2\pi}{a}$) is negligible with strain. Therefore, with a sufficient tensile strain, the energy of the valence band at $\Gamma$ can be reduced to an extent so that it is lower than the energy at $K = 0.05 \cdot \frac{2\pi}{a}$. The reason of the different band energy shifts at $\Gamma$ and other K vectors with strain is due to their different $s$, $p$, $d$ orbital projections, where the significant portion of $p_z/d_z^2$ exists at Γ.  The $p_z/d_z^2$ orbital suggests that the nodal surfaces of the positive and negative values of the wave-function are perpendicular to the axis of the wire [44, 45]. Under a uniaxial strain, the distance between the nodal surfaces will be modulated. which leads to a change of kinetic energy associated with the electron transportation between atoms [45]. This is in contrast with the case at other K vectors. (See the above section for detailed discussion).

The strain effects on the effective masses of the electron and hole are reported in Figure 7. The variation of the effective mass of the electron with strain for three wires with a diameter of 2.5 nm, 3.0 nm and 3.7 nm are minimal (see Figure 7(a) and 7(b)). For the wire with a diameter of 4.7 nm, the effective mass of the electron slightly increases with the compressive strain. In contract, the change in the effective mass of the hole is dramatic, as shown in Figure 7(c) and 7(d). Taking the 2.5 nm Ge-core/Si-shell wire as an example, the effective mass of the hole decreases from 0.21 $m_e$ (no strain) to 0.15 $m_e$ (-2.2% strain), which is indicated by the dispersion relations in Figure 6(a) shown by the curves with solid dots and squares. In the case of the 3.0 nm Ge-core/Si-shell wire, the effective mass of the hole decreases slightly from 0.166 $m_e$ without strain to 0.145 $m_e$ at -2.3% strain (decreased by 13%), while it dramatically increases to 0.728 $m_e$ at 0.7% tensile strain (increased ~ 300%). When a +1.7% tensile strain is applied, the VBE of the wire shifts from Γ to the K vector at $0.04 \cdot \frac{2\pi}{a}$. The effective mass of the hole was calculated to be 0.453 $m_e$ by a parabolic fitting the dispersion relation near this new K vector. Similar behavior is also observed for other wires, as shown in Figure 7(c) and 7(d). It is concluded that the effective mass of the hole in the Si/Ge core-shell nanowires is much more sensitive to strain, compared to that of the electron. This is implied in Fig. 6(a), which shows that the curvature of the conduction band around $\Gamma$ is not sensitive to strain. However, the curvature of the valence band is dramatically changed with strain.

## 4. Conclusion

In summary, it was found from our calculations that (1) the band structures of Si/Ge core-shell nanowires along the [110] direction can be tuned by external uniaxial strain; (2) the band gap of the wires varies with strain; (3) strain modulates the effective masses of the electron and hole in a different manner: strain has a minimal effect in changing the effective mass of the electron, while it can dramatically modify the effective mass of the hole.

## Acknowledgement

This work is supported by the Research Initiative Fund from Arizona State University (ASU) to Peng. The authors thank the following for providing computational resources: ASU Fulton High Performance Computing Initiative (Cluster Saguaro), National Center for Supercomputing Applications and Pittsburgh Supercomputing Center. Andrew Copple is acknowledged for review the manuscript.

**Table captions**

**Table 1** A list of the studied Si/Ge core-shell nanowires along the [110] direction in present work. D is the diameter of a wire; $N_{core}$ and $N_{shell}$ are the numbers of Si/Ge atoms in the core and shell of a given wire; $N_H$ is the number of H atoms needed to saturate the surface dangling bonds.

**Table 2** The optimized axial lattice constants, DFT predicted band gap and effective masses of the electron and the hole in Si/Ge core-shell nanowires along the [110] direction.

**Table 3** Projections of the valence band (VB) and the conduction band (CB) onto the *s*, *p*, and *d* orbits in the 2.5 nm wires.

**Figure captions**

**Figure 1**  Snapshots of the Si/Ge core-shell nanowires with a varying size viewed from the cross-section. Blue dots are shell atoms (Ge or Si), yellow are core atoms (Si or Ge), white are H atoms.

**Figure 2**  The band structures of Si/Ge core-shell nanowires at different diameters [(a)-(h)] and the band structure of 2.5 nm Ge-core/Si-shell nanowire under different value of uniaxial strain [(i)-(m)].

**Figure 3**  The iso-value surfaces of the charge density of the valance/conduction band edges in the Si/Ge core-shell nanowires. The value of charge density employed is 0.0004 e/(a.u.)$^3$. The yellow, blue and white dots represent Si, Ge, and H atoms, respectively.

**Figure 4**  The variation of the band gap in the Si/Ge core-shell wires with uniaxial strain. $\Delta E_g$ is defined as the difference of band gap with and without strain. Positive strain refers to uniaxial tensile strain while negative strain corresponds to its compression. (The data points indicated by an arrow are for indirect band gaps.)

**Figure 5**  The energy variation of the CBE and VBE with the external strain in the Si-core/Ge-shell nanowire with a diameter of 2.5 nm. The vertical axis is the energy difference of the CBE/VBE between the strained and relaxed wires.

**Figure 6**  The energy dispersion relation of the conduction/valence band of the Ge-core/Si-shell wire with a diameter of 2.5 nm near $\Gamma$. The rectangle enclosed curve under +1.8% strain was enlarged in (b), indicating an indirect band gap with the valence band edge located at the state $v_1$ rather than $v_0$.

**Figure 7** The effective masses of the electron (top) and hole (bottom) are plotted as a function of uniaxial strain for the Si/Ge core-shell nanowires at different size. It shows that the effective mass of the electron changes mildly with strain. However, the effective mass of the hole reduces under compression, while enhanced dramatically with tensile strain.

| D (nm) | $N_{core}$ | $N_{shell}$ | $N_H$ |
|---|---|---|---|
| 2.5 | 30 | 46 | 28 |
| 3.0 | 30 | 80 | 32 |
| 3.7 | 30 | 142 | 44 |
| 4.7 | 30 | 246 | 52 |

**Table 1**

| | Diameter (nm) | Optimized Axial Lattice (nm) | $E_g$ (eV) | $m_e^*$ | $m_h^*$ |
|---|---|---|---|---|---|
| Si-core /Ge-shell | 2.5 | 0.3917 | 0.54 | 0.13 | 0.16 |
| | 3.0 | 0.3945 | 0.29 | 0.13 | 0.21 |
| | 3.7 | 0.3965 | 0.18 | 0.14 | 0.32 |
| | 4.7 | 0.3944 | 0.13 | 0.14 | 0.26 |
| Ge-core /Si-shell | 2.5 | 0.3985 | 0.58 | 0.14 | 0.21 |
| | 3.0 | 0.3950 | 0.31 | 0.13 | 0.17 |
| | 3.7 | 0.3931 | 0.23 | 0.14 | 0.74 |
| | 4.7 | 0.3900 | 0.18 | 0.14 | 0.36 |

**Table 2**

| wire | state | $|s\rangle$ | $|p_x\rangle$ | $|p_y\rangle$ | $|p_z\rangle$ | $|d_{xy}\rangle$ | $|d_{yz}\rangle$ | $|d_{z^2}\rangle$ | $|d_{xz}\rangle$ | $|d_{x^2}\rangle$ |
|---|---|---|---|---|---|---|---|---|---|---|
| 2.5 nm Si-core /Ge-shell | VB at Γ | 0.014 | 0.000 | 0.000 | 0.995 | 0.000 | 0.099 | 0.000 | 0.002 | 0.000 |
| | VB at 0.2·2π/a | 0.045 | 0.967 | 0.212 | 0.098 | 0.095 | 0.008 | 0.011 | 0.003 | 0.005 |
| | CB at Γ | 0.596 | 0.022 | 0.585 | 0.000 | 0.000 | 0.000 | 0.547 | 0.000 | 0.044 |
| | CB at 0.2·2π/a | 0.603 | 0.155 | 0.532 | 0.054 | 0.012 | 0.006 | 0.066 | 0.006 | 0.567 |
| 2.5 nm Ge-core /Si-shell | VB at Γ | 0.008 | 0.000 | 0.000 | 0.995 | 0.000 | 0.097 | 0.000 | 0.002 | 0.000 |
| | VB at 0.2·2π/a | 0.059 | 0.991 | 0.073 | 0.018 | 0.089 | 0.002 | 0.005 | 0.005 | 0.002 |
| | CB at Γ | 0.588 | 0.022 | 0.669 | 0.000 | 0.005 | 0.000 | 0.452 | 0.000 | 0.044 |
| | CB at 0.2·2π/a | 0.646 | 0.152 | 0.545 | 0.108 | 0.038 | 0.025 | 0.063 | 0.013 | 0.494 |

**Table 3**

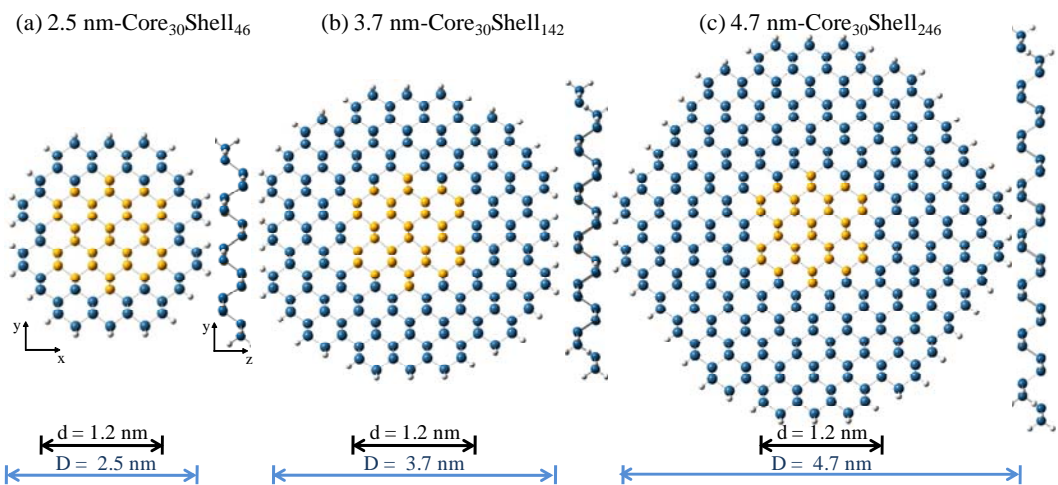

**Figure 1**

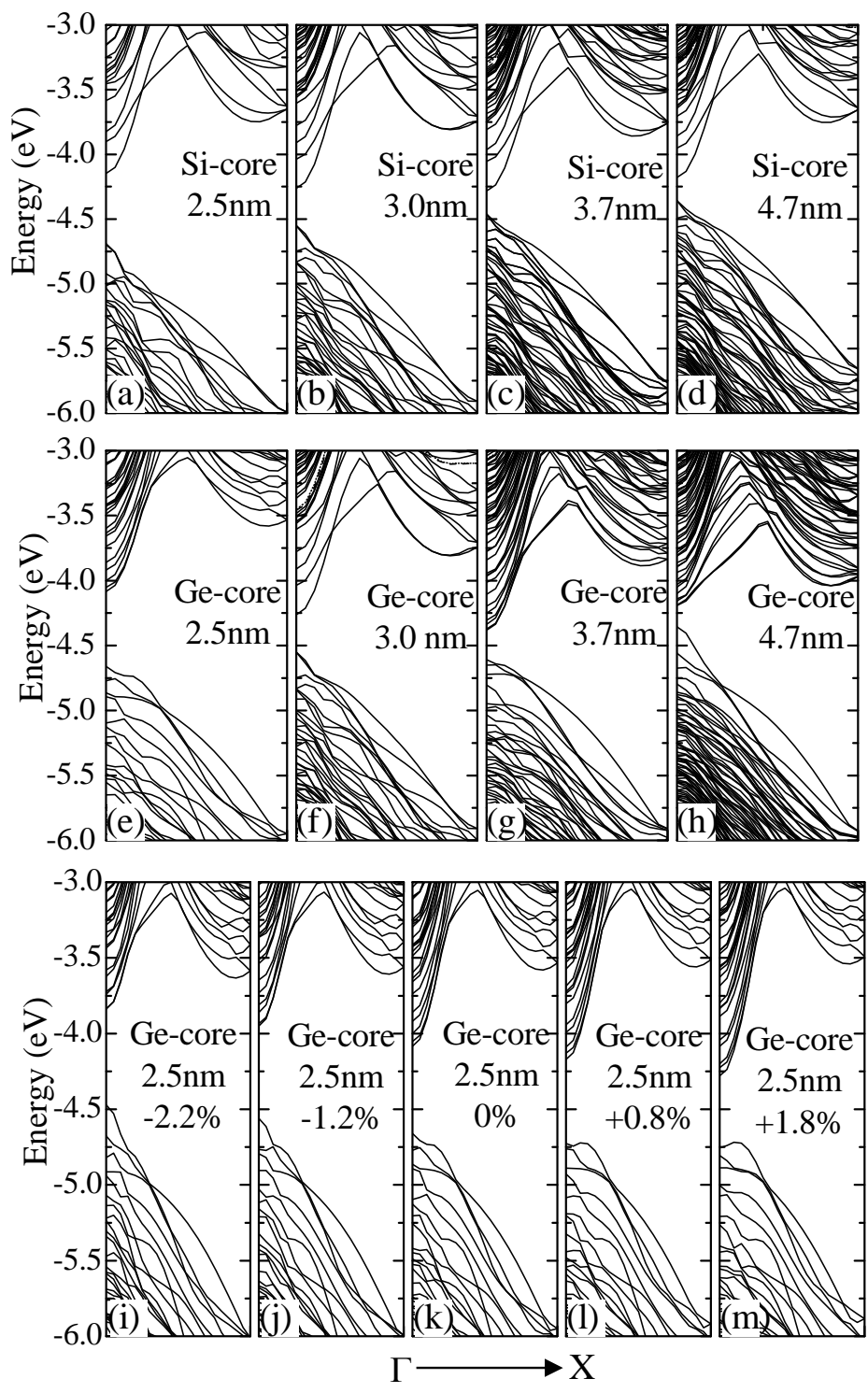

**Figure 2**

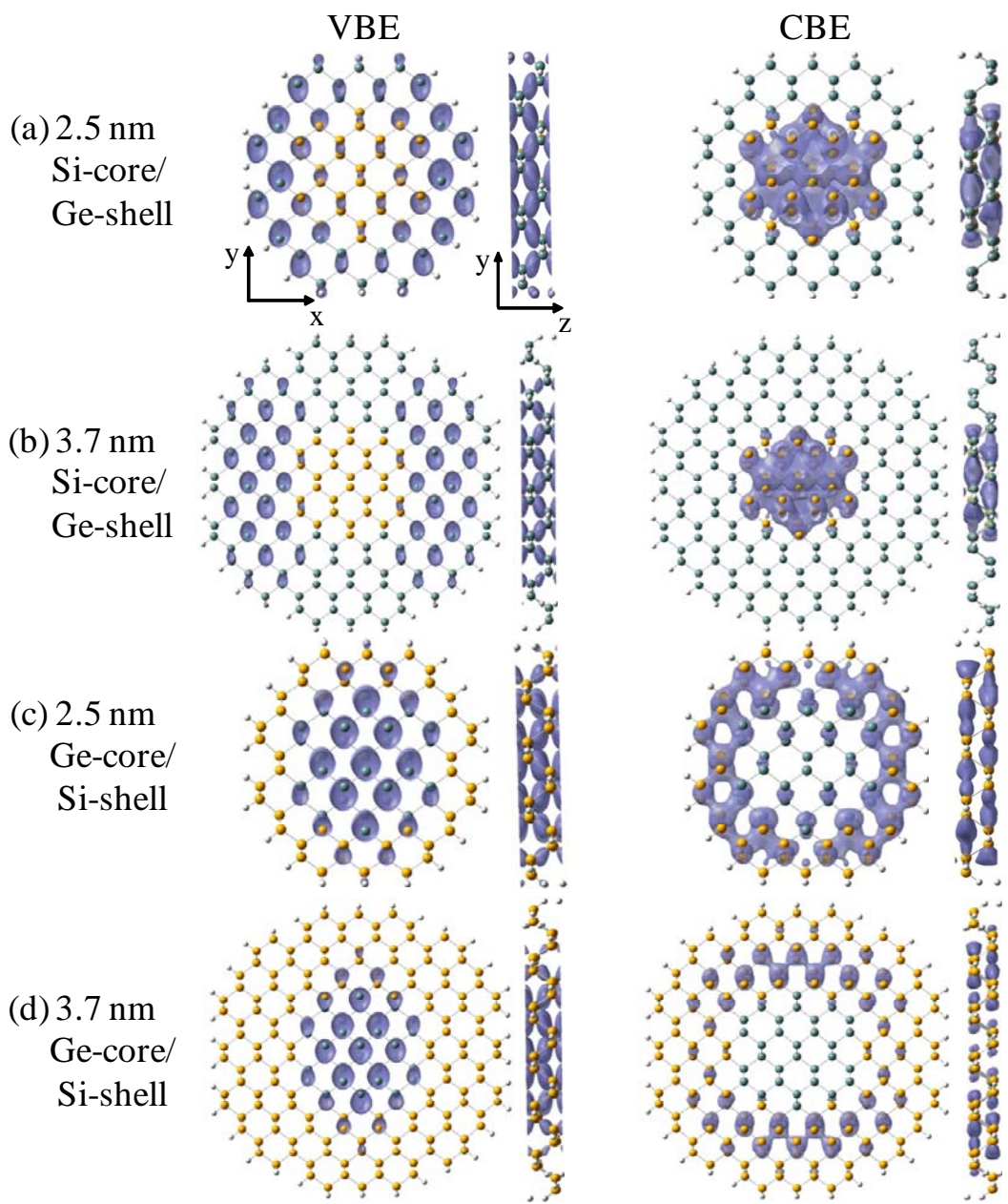

**Figure 3**

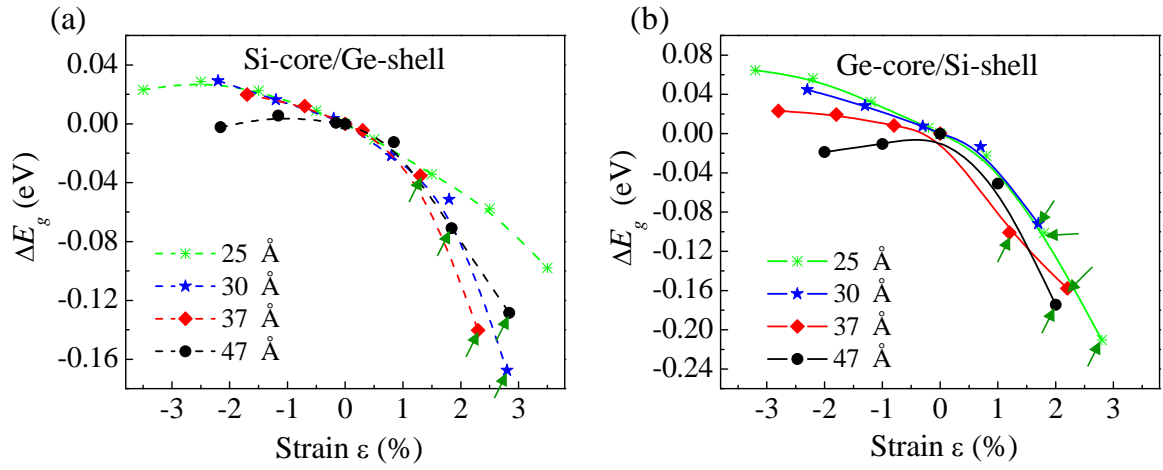

**Figure 4**

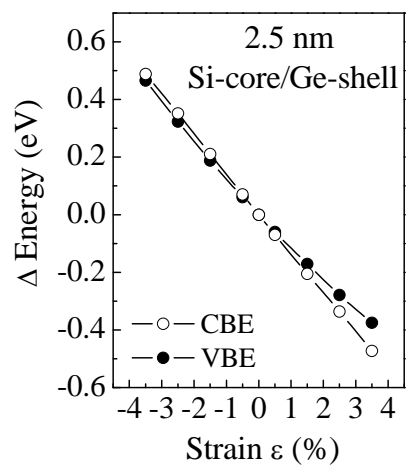

**Figure 5**

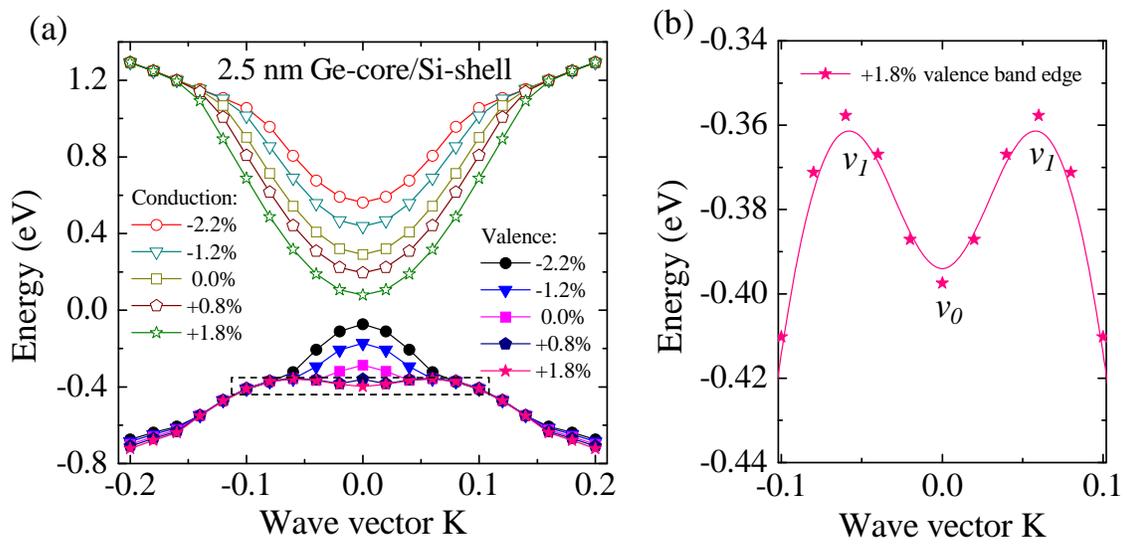

**Figure 6**

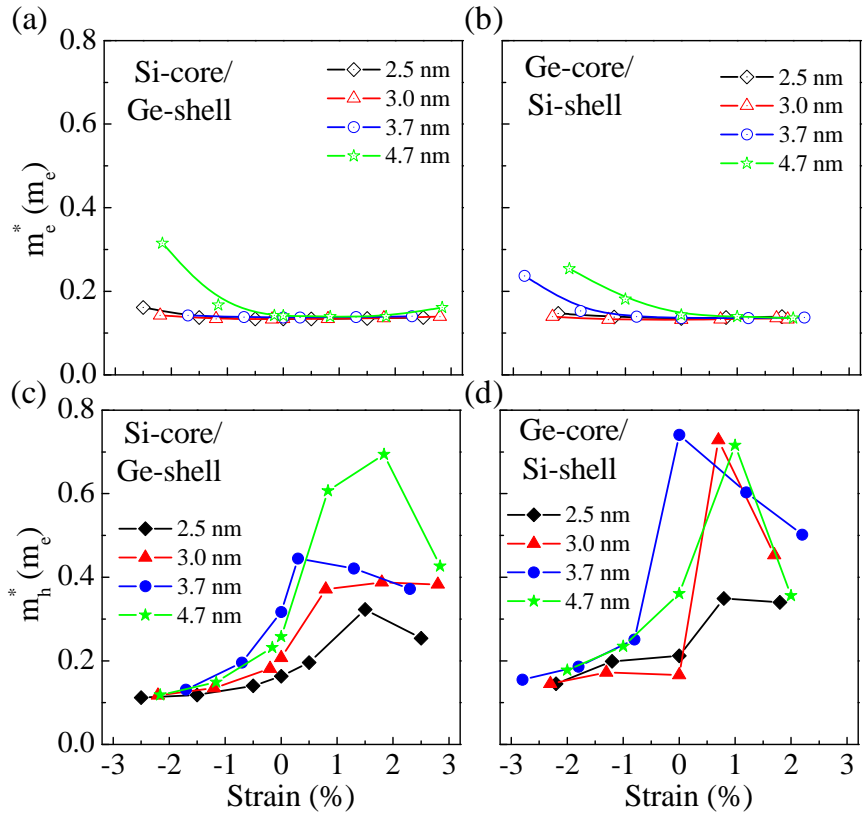

**Figure 7**